\documentclass[final]{ias2}

\usepackage{graphicx} 
\usepackage{multirow}
\usepackage{array} 

\usepackage{hyperref} 
\usepackage{bbm} 
\def\ii{{\rm i}}
\def\ket#1{|#1\rangle}
\def\bra#1{\langle#1|}

\begin{document}


\title{Quantum 
transport in 1d systems 
via a master equation approach: numerics and an exact solution}

\author[sin,ain]{Marko \v Znidari\v c} 
\address[sin]{
Instituto de Ciencias F\' isicas, Universidad Nacional Aut\' onoma de M\' exico, C.~P.~62132, Cuernavaca, Morelos, Mexico
}
\address[ain]{
Department of Physics, Faculty of Mathematics and Physics, University of Ljubljana, Jadranska 19, SI-1000 Ljubljana, Slovenia
}

\begin{abstract}
We discuss recent findings about properties of quantum nonequilibrium steady states. In particular we focus on transport properties. It is shown that the time dependent density matrix renormalization method can be used successfully to find a stationary solution of Lindblad master equation. Furthermore, for a specific model an exact solution is presented.
\end{abstract}



\maketitle


\section{Introduction}
Phenomenological laws of transport have been established long ago. In 1807 Joseph Fourier submitted his manuscript "The\' orie de la Propagation de la Chaleur dans les Solides'' to the French Academy. In it he presented several groundbreaking advances. He described heat conduction by a partial differential equation, equating time derivative of the temperature with its laplacian, whereas previous attempts mostly focused on describing conduction between two discrete objects. By using a continuum description Fourier could use a linear equation (instead of some complicated ``n-body''-like equation) for which he used a superposition principle. He solved the equation by separating variables and using a trigonometric series for the solution -- what is today known as Fourier's series. Fourier's law, stating that the heat current is proportional to the temperature gradient, lead subsequently to other phenomenological transport laws, like Ohm's or Ficks's law, all having the same form: current is proportional to the gradient of the driving field. According to standard Academy's practice, a referee committee consisting of Laplace, Lagrange, Monge, and Lacroix has been appointed in order to judge the suitability of the Fourier's manuscript. Laplace and Lagrange did not receive the manuscript very well though. In particular, they were not in favor of the trigonometric series that Fourier used in his solution and the manuscript has never been published by the Academy. Eventually, in 1822 Fourier published it by himself as "The\' orie Analytique de la Chaleur''. For more on rich history behind the Fourier's law see~\cite{Narasimhan:99,Narasimhan:09}.

More than 200 years later microscopic origin of Fourier's law is still not very well understood~\cite{Bonetto:01}. Given a systems, for instance described by a Hamiltonian, one can ask whether this system obeys Fourier's law or not. It is by no means granted that the Fourier's law will hold. For instance, taking a system of coupled harmonic oscillators Fourier's law is obviously violated. This happens because there are normal modes -- the phonons -- which do not interact and propagate in a ballistic fashion. In such a system the current is proportional to the temperature difference and not its gradient. Therefore, at fixed driving the current is in a ballistic system independent of the system's size $n$, $j \sim n^0$, whereas it scales as $j \sim 1/n$ if Fourier's law holds. Grossly speaking, the question therefore is, how does the current scale with the system size? If it is independent of $n$ we say that the system is a ballistic conductor, if it decreases as $\sim 1/n$ we say that the system is a normal conductor (sometimes, by a small abuse of language, also diffusive, because diffusion implies Fourier's law; note though that Fourier's law does not necessarily imply diffusion).

Based on the above example of a harmonic oscillator chain one could argue that all integrable systems would be ballistic. However, thing are not that simple. For instance, taking one of the simplest quantum models, a one-dimensional Heisenberg model,
\begin{equation}
H=J\sum_{j=1}^{n-1} \sigma_j^{\rm x} \sigma_{j+1}^{\rm x} +\sigma_j^{\rm y} \sigma_{j+1}^{\rm y}+ \Delta \sigma_j^{\rm z} \sigma_{j+1}^{\rm z},
\label{eq:Heis}
\end{equation} 
spin conduction for $\Delta \ge 1$ is still debatable, see references in e.~g.~\cite{Meisner:07}. Heisenberg model has been introduced by W.~Heisenberg in 1928~\cite{Heisenberg:28} as a model system to explain the ferromagnetism. Existing theories at the time, based on a direct interaction between magnetic moments, could not explain very high transition temperatures in ferromagnets. Heisenberg used the symmetry of wave functions to derive the so-called exchange interaction, resulting in an effective Heisenberg model. Heisenberg model in 1d is integrable by Bethe ansatz~\cite{Bethe:31}. Despite integrability the evaluation of observables is rather involved. Calculating time-dependent correlation functions, which could be used in the Kubo formula for conductivity, is at present not possible. Even for such simple system one therefore has to use either numerical computations of approximate techniques. If one uses Jordan-Wigner transformation~\cite{Jordan:28} between spin-1/2 particles and spinless fermions one can rewrite 1d Heisenberg model as a system of interacting fermions. It is therefore one of the simples so-called strongly-correlated systems, where many-body effects play a role. Although Heisenberg model has been initially thought of as being a simple toy model, today we know that it is realized to a very high degree of accuracy in many spin-chain materials, for a review see e.g~\cite{Sologubenko:07}.

The simplest transport situation is that of a stationary state, in which expectations do not change in time. Besides its relevance for transport such a stationary nonequilibrium state is one of the simplest nonequilibrium situations that one can study. Physics of nonequilibrium systems is much less understood that that of equilibrium. Detailed balance, which greatly simplifies equilibrium discussion, does not hold out of equilibrium. In stationary nonequilibrium situation only the total probability flow out of a state has to be zero and not along each connection individually as in equilibrium. Furthermore, there are very few exactly solvable nonequilibrium models from which we could learn about generic properties of nonequilibrium steady states (NESS). This is particularly true for quantum systems, whereas in classical physics there is a notable class of exactly solvable nonequilibrium stochastic lattice gasses~\cite{classical}. In quantum domain there are basically to ways in which one can study stationary nonequilibrium phenomena: (i) one can consider Hamiltonian evolution of a combined central-system + environment system and from its exact solution deduce the dynamics of the central system, or (ii) one can trace out environmental degrees of freedom before solving the system, deriving a master equation describing the evolution of the central system only. To use the first approach one obviously has to know how to exactly solve the total system. This is possible only in few instances, for instance for an XY spin chain~\cite{XYinf} which is equivalent to free fermions, or for the so-called star system~\cite{Breuer:04}. The second approach, in terms of a master equation, also allows for an explicit solution if the superoperator can be written as a quadratic function of fermionic operators~\cite{3rd}. 

In the present work we shall discuss quantum transport, and nonequilibrium physics in general, by studying solutions of master equation in a nonequilibrium setting. We are going to be in particular interested in the NESS as they are the simplest nonequilibrium states. Master equation that we shall use is the Lindblad equation~\cite{Lindblad,book},
\begin{equation}
\frac{{\rm d}}{{\rm d}t}{\rho}=\ii [ \rho,H ]+ {\cal L}^{\rm dis}(\rho)={\cal L}(\rho),
\label{eq:Lin}
\end{equation}
where the dissipative linear operator ${\cal L}^{\rm dis}$ is expressed in terms of Lindblad operators $L_k$,
\begin{equation}
{\cal L}^{\rm dis}(\rho)=\sum_k \left( [ L_k \rho,L_k^\dagger ]+[ L_k,\rho L_k^{\dagger} ] \right).
\end{equation}
Derivation of the Lindblad equation from the first principles, that is from a particular system-bath Hamiltonian, involves a number of approximations~\cite{book}. Note though that Lindblad equation describes the most general completely positive trace-preserving dynamical semigroup. Notwithstanding this, we know that there are evolutions that can not be described by a Lindblad equation. Our approach here to these issues is pragmatical. We shall not discuss approximations involved, neither if they are fulfilled, nor if Lindblad description is valid. The ``philosophy'' is that rather than study some more exact, but also more complicated, description about whose solutions we are not able to say much, we shall choose simple Lindblad equation with local dissipative terms which is amenable to solution. Our goal is to find solutions, even if they are for a simplified model system. After all, one of the reasons for the success of physics is that often simple model systems display generic behavior. In view of the lack of knowledge about quantum nonequilibrium system we fell that description with local Lindblad equation is the way to proceed at this early state of investigations.    

We shall use two approaches to find the NESS of Lindblad equation. In the second part we shall present a model for which an exact solution is possible for any system size $n$. For systems which are not exactly solvable one has to either use approximations~\cite{approx} or resort to numerics. Due to exponential scaling of computational complexity with $n$, until very recently, numerics has been limited to chains of fairly small size of less than $\approx 20$ spins~\cite{master,Michel:08} from which it is hard to conclude about properties in the thermodynamic limit which is of especial interest to us. Recently however, a time dependent density matrix renormalization group method (tDMRG) has been presented~\cite{vidal}, which allows to study much larger systems. tDMRG can be used to simulate unitary pure state evolution as well as evolution described by master equation. Transient nonequilibrium studies of non-stationary pure states using tDMRG are quite abundant, an incomplete list includes for instance studies of transport via the spreading of inhomogeneities~\cite{packets} or $I-V$ characteristics~\cite{IV}. On the other hand there are much fewer studies of time evolution within a master equation approach. In our view master equation formulation has certain theoretical and practical advantages over pure state evolution, see discussion below. In the first part we shall therefore present tDMRG method and its application on finding NESS and calculating transport properties. Both methods, analytical solution and numerical investigation, enabled us to study some interesting phenomena which are not found in equilibrium systems, like nonequilibrium phase transitions and a ubiquitous presence of long-range order. 

\section{Numerical study}
\begin{figure*}
\begin{center}
\includegraphics[width=0.7\textwidth]{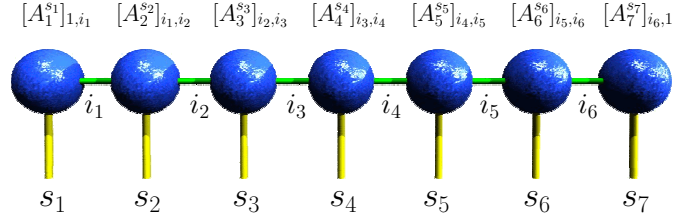}
\caption{Schematic representation of a MPO (\ref{eq:MPO}) with $7$ sites. Each site is described by matrices (balls) having a local basis-label ($s_j$ -- yellow vertical lines) and two matrix indices (horizontal lines).}
\label{fig:mps}
\end{center}
\end{figure*}

We shall explain the working of tDMRG algorithm to calculate NESS on a case study of spin transport in the Heisenberg model (\ref{eq:Heis}) with the anisotropy $\Delta=1.5$ ($J=1$) at infinite temperature reported in~\cite{JSTAT09}. First, the dissipative Lindblad part ${\cal L}^{\rm dis}$ will act only on the 1st and the last spin. Each of these two parts has two Lindblad operators, one being proportional to $\sigma^+=(\sigma^{\rm x}+\ii \sigma^{\rm y})/2$ and the other to $\sigma^-=(\sigma^{\rm x}-\ii \sigma^{\rm y})/2$. If the coefficients in front of these two operators are different, they will try to induce a net nonzero magnetization on the corresponding spin. They can be expressed in terms of single bath parameter $\mu$ playing the role of driving potential~\cite{JSTAT09} (in fact, $\mu=\beta \phi$, if $\beta$ is the inverse temperature and $\phi$ chemical potential).
 Having different $\mu$ at the left and right ends, in our case we take $\mu_{\rm left}=0.02$ and $\mu_{\rm right}=-0.02$, will induce a nonequilibrium driving. 
\begin{figure*}
\begin{center}
\includegraphics[width=0.9\textwidth]{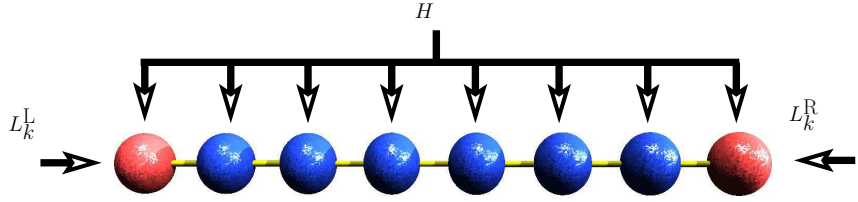}
\caption{Schematic representation of our Lindblad equation (\ref{eq:Lin}). In addition to unitary part, dissipative part due to baths acts on the first and the last spin.}
\label{fig:chainL}
\end{center}
\end{figure*}
Starting with some initial density matrix $\rho(0)$ we want to calculate $\rho(t)$ at time $t$ being a solution of the Lindblad equation (\ref{eq:Lin}). The asymptotic stationary state $\rho(t \to \infty)$ is then the sought-for NESS. To calculate $\rho(t)$ as efficiently as possible we write it in terms of the so-called matrix product operator (MPO) representation,
\begin{equation}
\rho=\frac{1}{2^n}\sum_{s_j} \bra{1} A_1^{(s_1)} A_2^{(s_2)} \cdots A_n^{(s_n)}\ket{1}\,\, \sigma_1^{s_1} \sigma_2^{s_2} \cdots \sigma_n^{s_n}.
\label{eq:MPO}
\end{equation}
For each site $j$ we have four matrices $A^{(s_j)}$, each of dimension $D \times D$, for schematic representation see fig.~\ref{fig:mps}.
The advantage of a MPO representation lies in the fact that if the state $\rho$ has small entanglement then the matrix size $D$ can be small and independent of $n$ so that the number of parameters needed to describe the state scales as $\sim nD^2$, whereas it would scale exponentially in $n$ in the worst case of large entanglement because the matrices would have to be exponentially large. Interesting fact is that some states, like ground states in 1d, do have such small entanglement. For more on matrix product states see the review~\cite{Murg}.

\begin{figure*}[ht!]
\begin{center}
\includegraphics[width=0.6\textwidth]{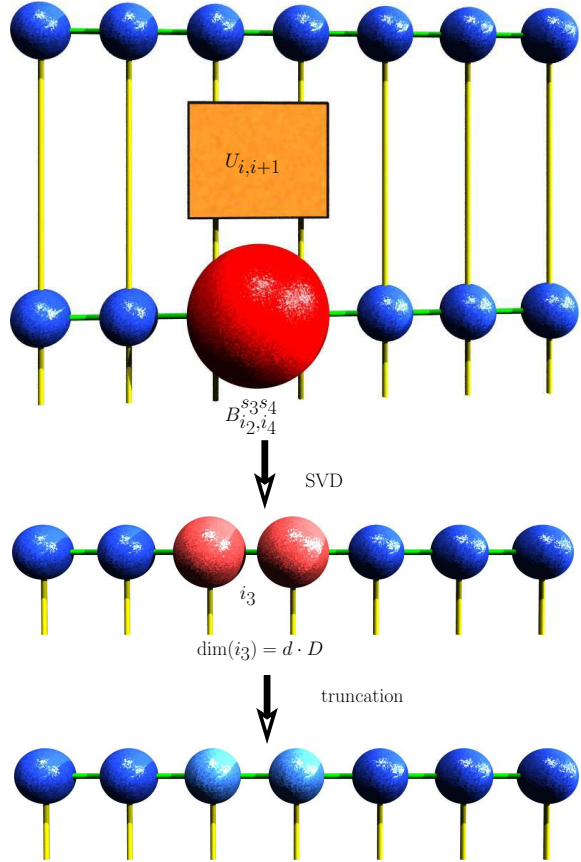}
\caption{Schematic working of one short time-step transformation $\exp{({\cal L}_{j,j+1} \delta t)}$ acting on $j$-th and $(j+1)$-th spins of an MPO ansatz. Algorithm runs from top to bottom.}
\label{fig:Umps}
\end{center}
\end{figure*}
Time evolution according to the Lindblad master equation now proceeds in small time steps of length $\delta t$. The formal solution $\rho(t)=\exp{({\cal L} t)} \rho(0)$ is decomposed using Trotter-Suzuki formula into small time-step operations on nearest neighbor spins. Decomposition into operations acting only on nearest-neighbor spins is crucial and connected to the fact that matrix indices connect only neighboring spins, see fig.~\ref{fig:mps}. Performing nearest-neighbor transformation on the MPO ansatz transforms the two neighboring matrices into one larger one. To restore a MPO form a singular value decomposition is performed, rewriting the state $\rho(t+\delta t)$ in a MPO form, but with the two matrices being of a larger size. If they have been of size $D$ before the application of the transformation, they are now of size $4\cdot D$. To prevent an exponential increase of matrix size with time we have to truncate them, usually back to the original dimension $D$. Doing that we of course make some truncation error. The size of it depends on the discarded singular values, which are in turn equal to the squares of Schmidt coefficients, i.e. the entanglement. One short time step of tDMRG algorithm, acting on nearest neighbors, is pictorially described in fig.~\ref{fig:Umps}. Technical details can be found in original references~\cite{vidal}, whereas our implementation for the Lindblad equation is described in~\cite{JSTAT09} and~\cite{NJP10}. Crucial question for the efficiency of the algorithm now is how big must the matrices be for the truncation error to be bearable? For unitary time evolution (that is without a dissipative part) entanglement for generic (chaotic) systems always grows with time and with it also the necessary dimension $D$. tDMRG method for conservative systems is therefore not efficient because it breaks down after relatively short time~\cite{PRE07}. It seems that things are not that bad for open systems though. As a rule dissipation namely decreases the entanglement. Although no systematic study has been performed to date it seems that often one can sufficiently accurately describe NESS with matrices of the order $D \sim 100$ for chain lengths of order $n \sim 100$. Using tDMRG on the Lindblad master equation one can therefore calculate NESS for significantly larger systems than with other methods. 

\begin{figure}[ht]
\begin{center}
\includegraphics[width=0.7\columnwidth]{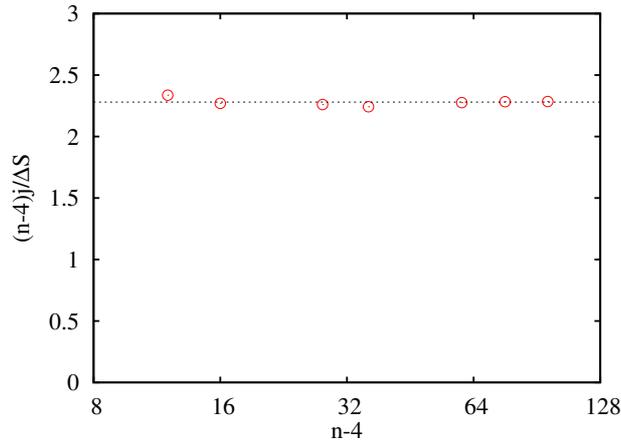}
\caption{Normal spin transport in Heisenberg model at $\Delta=1.5$; we show a scaled current divided by the magnetization difference. Horizontal line shows the value of the transport coefficient $\kappa\approx 2.3$. Data from~\cite{JSTAT09}.}
\label{fig:j}
\end{center}
\end{figure}

For our spin conduction example we could calculate spin current $j_k=2(\sigma_k^{\rm x} \sigma_{k+1}^{\rm y}-\sigma_k^{\rm y} \sigma_{k+1}^{\rm x})$ in the NESS for systems of various sizes. From the figure~\ref{fig:j} one can see that the current indeed scales as $j_k \sim 1/n$. At infinite temperature gaped Heisenberg model therefore seems to display normal spin transport. Defining transport coefficient $\kappa$ in terms of a gradient of magnetization, $j \asymp \kappa (z_n-z_1)/n$, we get $\kappa \approx 2.3$. Approximately the same value of the transport coefficient is obtained also from the equilibrium correlation function using Kubo formula~\cite{Steinigeweg:09} and qualitatively agrees with findings in~\cite{Samir:04,Michel:08}.
Because we have defined $\kappa$ in terms of magnetization, whose conjugate variable is $\mu=\beta \phi$, $\kappa$ is actually the diffusion coefficient. Spin conductivity $\sigma$, which is defined in terms of the gradient in the potential $\phi$, is then simply $\sigma=\beta \kappa$. 

For the end let us list some advantages and disadvantages of using tDMRG to calculate NESS. Among advantages are: (i) dissipative part of Lindblad equation seems to decrease entanglement significantly as compared to unitary evolution. Precise scaling is not known but simulating systems of $n \approx 100$ spins is feasible with modest computational resources; (ii) For the efficiency of calculating NESS the convergence time after which we get the asymptotic state $\rho(t \to \infty)$ is also relevant. The convergence time is given by the inverse gap of the superoperator ${\cal L}$ (\ref{eq:Lin}). The scaling of the gap has been studied in some exactly solvable systems~\cite{3rd,exact} and has been found to be always polynomial in $n$; (iii) Solving for NESS of a master equation one is actually as close as possible to an actual experimental situation. On the down side we can mention: (i) For stronger driving $\mu$ the entanglement increases and the method is not so efficient. This in particular makes it harder to study~\cite{NDC} some interesting phenomena that occur only at strong driving, like negative differential conductivity. (ii) For density matrices tDMRG is more computationally intensive than for pure states (provided we have the same $D$). (iii) It works efficiently only in 1d. (iv) No theorems exist about the entanglement of NESS states that would guarantee the efficiency of tDMRG as opposed to ground-state situations. (v) With local Lindblad operators it seems that it is in general difficult to simulate baths at very low temperatures~\cite{PRE10}. In this regime it looks that pure state evolution using tDMRG~\cite{packets} might have certain advantages.   

\section{An exact solution}
An exact solution for NESS is possible in systems whose superoperator ${\cal L}$ is a quadratic function of fermionic operators~\cite{3rd}. Recently an exact solution~\cite{exact} has been found also for a system that is not quadratic in fermions. The unitary bulk part is described by XX Hamiltonian,
\begin{equation}
H=\sum_{j=1}^{n-1} (\sigma_j^{\rm x} \sigma_{j+1}^{\rm x} +\sigma_j^{\rm y} \sigma_{j+1}^{\rm y}).
\label{eq:H}  
\end{equation}
Coupling to the baths at boundary two spins is the same is in the example in previous section. At each side we have two Lindblad operators, 
\begin{equation}
L^{\rm L}_1=\sqrt{(1-\mu)}\,\sigma^+_1,\qquad L^{\rm L}_2=\sqrt{(1+\mu)}\, \sigma^-_1,
\label{eq:Lbath}
\end{equation}
at the left end, while at the right end we have 
\begin{equation}
L^{\rm R}_1=\sqrt{(1+\mu)}\,\sigma^+_n,\qquad L^{\rm R}_2=\sqrt{(1-\mu)}\, \sigma^-_n.
\end{equation}
With only these terms the model is actually quadratic in fermions and has been solved before~\cite{Karevski:09}. An interesting twist now comes if we add an additional dissipative part due to dephasing at each spin. That is, each spin site experiences an independent reservoir that causes the decay of off-diagonal matrix elements. Such dephasing can be described by a single Lindblad operator at each site,
\begin{equation}
L^{\rm deph}_j=\sqrt{\frac{\gamma}{2}}\sigma^{\rm z}_j.
\end{equation}
Dephasing makes the system nonquadratic because the dephasing part is quadratic in $\sigma^{\rm z}$, which is itself quadratic in fermionic operators. This renders the system more interesting, for instance, the NESS is not gaussian and the Wick theorem does not apply~\cite{exact}. Nevertheless, the NESS can be calculated exactly in a systematic way~\cite{exact}. First, one notices that the equilibrium stationary state, that is NESS for $\mu=0$, is a trivial identity,
\begin{equation}
\rho_{\rm eq}=\frac{1}{2^n}\mathbbm{1}.
\end{equation}
This can be interpreted as an infinite temperature state. For small driving $\mu$ the NESS will differ from $\rho_{\rm eq}$ by a small amount. The ansatz for the NESS is written as a formal series in $\mu$,
\begin{equation}
\rho=\frac{1}{2^n}\left( \mathbbm{1} + \mu R^{(1)}+\mu^2 R^{(2)}+\cdots \mu^r R^{(r)}+\cdots \right),
\label{eq:series}
\end{equation}
however, there is a crucial difference from the ordinary perturbation series. Due to algebraic structure, if one writes stationary equation ${\cal L}\rho=0$ satisfied by NESS, one can see that the unknown terms in the $p$-th order term $R^{(p)}$ depend only on coefficients in the lower order terms. Now because the zeroth order term is explicitly known, it is $\mathbbm{1}/2^n$, one can solve for terms order by order. Also, $p$-th order term is a sum of products of $p$ factors, each one being either a magnetization $\sigma_k^{\rm z}$, or a spin current $j_k$. This also means that connected $p$-point correlation functions are contained solely in $R^{(p)}$. First two orders are actually quite simple and we can write them in full. One gets
\begin{equation}
\mu R^{(1)}=\mu A + \mu B,\qquad \mu A= \sum_{j=1}^n a_j \sigma^{\rm z}_j,\quad \mu B=\frac{b}{2}\sum_{k=1}^{n-1}j_k.
\label{eq:1st}
\end{equation}
Magnetization profile given by $a_j$ is linear and equal to
\begin{equation}
a_j=-\mu+\mu\frac{2+2(j-1)\gamma+\delta_{j,1}-\delta_{j,n}}{2+(n-1)\gamma}, \qquad b=-\frac{\mu}{2+(n-1)\gamma}.
\end{equation}
We can see that the current scales as $1/n$ for large $n$. The system is therefore a normal conductor as long as the dephasing strength $\gamma$ is nonzero. 2nd order term can be written as
\begin{equation}
\mu^2 R^{(2)}=\frac{\mu^2}{2}\left(A B +B A \right)+\mu^2 C +\mu^2 D + \mu^2 F,
\end{equation}
\begin{eqnarray}
\mu^2 C&=&\sum_{j=1}^n \sum_{k=j+1}^n (C_{j,k}+a_j a_k) \sigma_j^{\rm z} \sigma_k^{\rm z},\nonumber \\
\mu^2 D&=&\sum_{j=1}^{n-2} \frac{d_j}{2} \left(\sum_{l=j+1}^{n-1}\sigma^{\rm z}_j j_l-\sum_{l=1}^{n-1-j}j_l \sigma^{\rm z}_{n+1-j} \right),\nonumber \\
\mu^2 F&=&\frac{f}{8}\sum_{k,l=1 \atop k\neq l}^{n-1} j_k j_l.
\end{eqnarray}
For large $n$ the dominant term in the 2nd order comes from $C_{j,k}$ whose exact expression is
\begin{equation}
C_{j,k}=-\frac{b^2}{(n-2)+2/\gamma}\left(u_j u_{n+1-k}+(n-1+2/\gamma)\delta_{j+1,k} \right),
\end{equation}
where $u_j=2+2(j-1)\gamma$. $C_{j,k}$ is equal to two-point connected correlation of magnetization. One can see that for $\gamma \neq 0$ there is a long-range order, i.e., at a fixed distance $|j-k|$ the correlation function goes to a constant plateau value of $\mu^2/n$ independent of $|j-k|$. The plateau is of purely nonequilibrium origin and decreases in the limit of $n \to \infty$. Similar long-range plateaus have been observed in quadratic nonequilibrium quantum systems~\cite{3rd}, see also~\cite{Temme}, leading to a conjecture~\cite{PRL10} that this a generic situation. 

\section{Discussion}
Solvable model presented in the last section is in a way minimal model that can display normal transport. Any nonequilibrium steady state must necessarily have a nonzero current and magnetization terms, plus any optional correlations between these two quantities (these correlations are perhaps necessary to have nonzero gradient). In our solution these are exactly the terms present. Namely, the NESS is composed only of products of local magnetizations and currents. The model exhibits some other interesting properties. The NESS can be written in terms of a MPO ansatz with matrices of small dimension $D=4$ that is independent of $n$~\cite{exact}. It also displays a nonequilibrium phase transition at $\gamma=0$, going from a state without long-range correlations to the one with.

On the numerical side, tDMRG shows to be a very useful method to explore nonequilibrium physics, especially in a stationary setting, where things are believed to be simpler than for instance in transient nonequilibrium situation (quantum quenches, for example). Considering that the field of nonequilibrium quantum physics is still rather unexplored it seems certain that many interesting things are still to be discovered.

\bibliographystyle{pramana}

\end{document}